**Full title:** Focal Loss Analysis of Peripapillary Nerve Fiber Layer Reflectance for Glaucoma Diagnosis

**Short title:** Focal loss analysis of NFL reflectance for glaucoma

**Author:** Ou Tan, PhD[1], Dongseok Choi[1,2], Aiyin Chen,[1] David S. Greenfield[3]; Brian A. Francis[4]; Rohit Varma[5]; Joel S. Schuman[6]; PhD, David Huang, MD, PhD[1], Advanced Imaging for Glaucoma Study Group

[1] Casey Eye Institute, Oregon Health & Science University

[2] OHSU-PSU School of Public Health, Oregon Health & Science University

[3] Bascom Palmer Eye Institute, University of Miami; [4] Doheny Eye Center and David Geffen School of Medicine at UCLA; [5] Southern California Eyecare and Vision research Institute; [6] New York University Langone Medical Center.

*Corresponding Author: Ou Tan, PhD

503-494-9436

Email: tano@ohsu.edu

**Funding Sources:** NIH grants R21 EY032146, R01 EY023285, and P30 EY010572, and an unrestricted grant from Research to Prevent Blindness to Casey Eye Institute.

**Commercial relationships:** OT: P; DC: None; AC: None; DH: F,R,I,P, DG: None; BF: None; RV: None; JS: None; PhD

Abstract

**Purpose**: To evaluate nerve fiber layer (NFL) reflectance for glaucoma diagnosis using a large dataset.

**Methods**: Participants were imaged with 4.9mm ONH scans using spectral-domain optical coherence tomography (OCT). The NFL reflectance map was reconstructed from 13 concentric rings of optic nerve head(ONH) scan, then processed by an azimuthal filter to reduce directional reflectance bias due to variation of beam incidence angle. The peripapillary thickness and reflectance maps were both divided into 96 superpixels. Low-reflectance and low-thickness superpixels were defined as values below the 5th percentile normative reference for that location. Focal reflectance loss was measured by summing loss, relative to the normal reference average, in low-reflectance superpixels. Focal thickness loss was calculated in a similar fashion. The area under receiving characteristic curve (AROC) was used to assess diagnostic accuracy.

**Results**: Fifty-three normal, 196 pre-perimetric, 132 early perimetric, and 59 moderate and advanced perimetric glaucoma participants were included from the Advanced Imaging for Glaucoma Study. Sixty-seven percent of glaucomatous reflectance maps showed characteristic contiguous wedge or diffuse defects. Focal NFL reflectance loss had significantly higher diagnostic accuracy than the best NFL thickness parameters (both map-based and profile-based): AROC 0.80 v. 0.75 (p<0.004) for distinguishing glaucoma eyes from healthy control eyes. The diagnostic sensitivity was also significantly higher at both 99% and 95% specificity operating points.

**Conclusions**: Focal NFL reflectance loss improved glaucoma diagnostic accuracy compared to the standard NFL thickness parameters.

**Translational Relevance:** The high diagnostic accuracy of NFL reflectance may make population-based screening feasible.

# 1 Introduction

Nerve fiber layer (NFL) thickness measured by optical coherence tomography (OCT) is widely used to diagnose and monitor glaucoma.[1-6] However, the diagnostic sensitivity is not sufficient to be used alone in screening due to the possibility of false negatives.[2, 7] The best single NFL thickness parameters have a sensitivity of only 7-30% for pre-perimetric glaucoma (PPG) and 20-60% for perimetric glaucoma (PG) at the 99% specificity diagnostic cutoff, which is needed in screening applications.[8-13] Studies showed better sensitivity of 55-85% for perimetric glaucoma by combining diagnostic parameters from several anatomic regions.[8, 10, 14-18] However, higher diagnostic accuracy is needed for application in population screening.

In our previous study, we found that focal loss analysis of NFL reflectance had significantly higher diagnostic accuracy for glaucoma than NFL thickness.[19] We developed an azimuthal filtering and applied focal NFL reflectance loss over a super grid to reduce the variability of NFL reflectance measurements. However, the sample size of the study was relatively small. The goal of the present paper is to validate the advantages of focal reflectance analysis with a large dataset from the Advanced Imaging for Glaucoma (AIG) study.[20]

Because the AIG study was conducted in an earlier period with a slower OCT system, the original NFL reflectance analysis[19] had to be adapted to the sparser sampling of the older OCT scans. Thus we do not expect the same level of diagnostic performance but hope to confirm the advantage of our method of NFL reflectance analysis over the standard thickness measurements.

# 2 Methods

## 2.1 Participants

Data from the Advanced Imaging for Glaucoma (AIG) study were analyzed. The AIG study was a bioengineering partnership and multi-site longitudinal prospective clinical study sponsored by the National Eye Institute (ClinicalTrials.gov identifier: NCT01314326). The study design and baseline participant characteristics have been reported previously,[20] and the Manual of Procedures is publically available online (www.AIGStudy.net). The study adhered to the Declaration of Helsinki and followed the Health Insurance Portability and Accountability Act of 1996 (HIPAA) privacy and security regulations. Written informed consent was obtained from all patients for participation in the study. Institutional review board approvals were obtained from all participating institutions.

In this secondary data analysis, participants who were normal (N), with pre-perimetric glaucoma (PPG), and perimetric glaucoma (PG) from the AIG study were analyzed. Half of the participants in the normal group of AIG study were used as normal references. The other half of normal participants were combined with glaucoma sub-groups to use as the testing dataset. The PG group was subdivided into early PG and Moderate+ PG groups based on visual field (VF), using mean deviation (MD) -6dB as a cutoff.

Each group's inclusion and exclusion criteria are:

- Normal group: VF tests within normal limits, IOP<21 mm Hg, and normal optic nerve on slit-lamp biomicroscopy. Both eyes of each participant are enrolled if they meet the criteria
- PPG group: glaucomatous optic neuropathy as evidenced by diffuse or localized thinning of the neuroretinal rim or NFL defect on fundus examination, but normal VF with pattern standard deviation (PSD, P >0.05) and glaucoma hemifield test (GHT) within normal limits,

- PG group: glaucomatous optic neuropathy as evidenced by diffuse or localized thinning of the neuroretinal rim or NFL defect on fundus examination, and corresponding repeatable VF defects with PSD (P < .05) or GHT outside normal limits, and
- Exclusion criteria common to all groups: best-corrected visual acuity (BCVA) worse than 20/40, evidence of retinal pathology, or history of keratorefractive surgery.

## 2.2 Data Acquisition

Participants were scanned with 27 kHz, 840 nm wavelength spectral-domain OCT systems (RTVue, Optovue, Inc., Fremont, CA, USA) from three clinic centers of the AIG study. Both eyes of each participant were scanned with the ONH scan twice or three times. The ONH scan was a 4.9 mm composite scan that centers on the disc and contains 13 concentric scans covering the peripapillary region (Figure 1A), and 12 radial scans covering the disc region. RTvue software provided RNFL thickness analysis(NFL thickness profile at D=3.4mm), the NFL thickness map, and ONH analysis for ONH scans.

The VF was assessed by standard automated perimetry on the Humphrey Field Analyzer (HFA II; Carl Zeiss Meditec, Inc., Dublin, CA, USA), using the Swedish Interactive Thresholding Algorithm 24-2.

## 2.3 NFL Reflectance Analysis

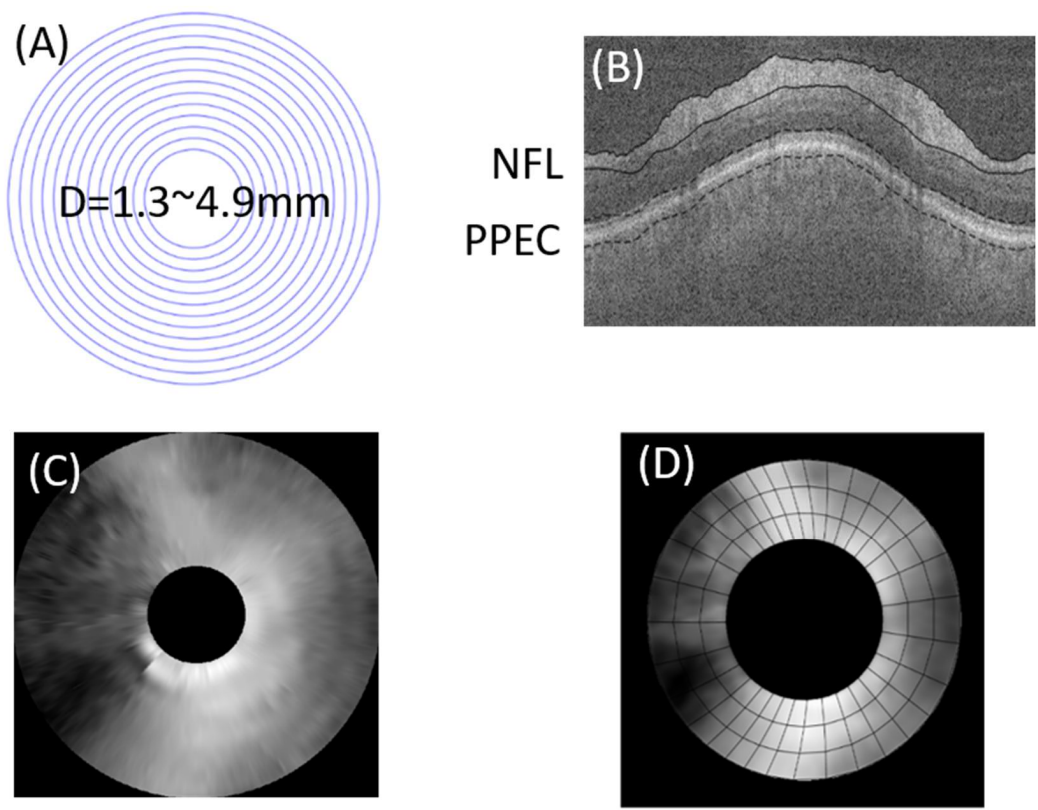

**Figure 1.** Calculation of the nerve fiber layer (NFL) reflectance map (4.9 mm) in a right eye with pre-perimetric glaucoma with an inferotemporal nerve fiber bundle defect. (A) ONH scan pattern with 13 concentric circular B-scans (B) Images were segmented to identify the topmost NFL and a reference layer called the photoreceptor-pigment epithelium complex (PPEC). (C) NFL/PPEC reflectance ratio map

reconstructed from 13 circular scans. (D) Reflectance map and super-pixel grid after re-centration and filtering Image segmentation

The image was segmented using the REVue Software (Version 6.12, Optovue). Scan with poor image quality or incorrect RNFL thickness detection were excluded from future analysis.[20-22] The graders reviewed and manually redrew the boundary if necessary. The boundaries included the inner limiting membrane (ILM), outer NFL, and retinal pigment epithelium (RPE) (Figure 1B). The OCT intensity image, disc center, boundary segmentation were exported for reflectance analysis.

### 2.3.1 NFL Reflectance Map

We adapted the NFL reflectance map calculation from the cubic scan to the ONH scan. In short, the reflectance was summed in the NFL band, from 7 pixels under ILM to the outer NFL boundary. Then the summation was normalized by the reflectance averaged in the photoreceptor and pigment epithelium complex (PPEC) band (Figure 1B). The reflectance values shadowed by the vessel were replaced with neighboring pixels to preserve continuity. The NFL reflectance map was interpolated from NFL reflectance ratio profiles from 13 rings (Figure 1C).

After re-centering the map to the disc center, we selected a 2.1 – 4.2 mm diameter analytic zone. The region outside the 4.2-mm diameter was excluded to avoid cropping artifacts from possible scan de-centration, while the region inside the 2.1mm diameter was excluded to mask the optic disc. Then we performed the azimuthal spatial frequency filtering by removing the first-degree angular component in the azimuthal dimension, as we have found that the azimuthal filter reduced the bias caused by the incident angle.[19]

The filtered NFL reflectance map was divided into superpixels (Fig. 1D). The superpixel grid contains 32 tracks parallel to the average nerve fiber trajectory map. Each track was evenly divided into three segments in the annulus between 2.1 and 4.2 mm from the center of the disc. The NFL reflectance in each superpixel was averaged.

### 2.3.2 Age, Gender, Axial Length and Clinical site Adjustment Using Linear Mixed Effects Model

Multiple linear regression based on the linear mixed-effects model[23, 24] was used to estimate the association between the normalized NFL reflectance in the normal reference and other variables (age, gene, axial length, and clinical site). The superpixel location was modeled as a random effect, while age, axial length, gender, and clinical site were tested as fixed effects. Only age, axial length, clinical site, and interactions were significant factors. We also tested variable sets of age, gender, axial length and ethnics, and only found age and axial length significant. Therefore, the NFL reflectance of superpixels was adjusted for age, axial length, and clinical site using the fitted regression model to the normal reference.

### 2.3.3 Focal Reflectance Loss Analysis

The 5% and 1% cutoff of reflectance values were estimated for each superpixel from the normal reference. Superpixels with adjusted reflectance below the 5% cutoff were considered "low-reflectance." Focal reflectance loss was the summation of reflectance deviation (the difference between the tested superpixel and the normal reference, adjusted for age, axial length, and clinic site) over the low-reflectance superpixels. Focal reflectance loss was then normalized by the total number of superpixels (n = 96).

### 2.3.4 Other Diagnostic Parameters

Besides the focal reflectance loss, we also calculated the overall average reflectance, which was the average of reflectance values in all superpixels.

NFL thickness measurements were obtained from both the RNFL thickness profile and NFL thickness map. Overall average NFL thickness and focal thickness loss were obtained from the NFL thickness profile at the 3.4-mm diameter circumpapillary circle.[22] Overall average NFL thickness and focal thickness loss were also obtained from the NFL thickness superpixel map (D=2.1~4.2mm) and following the same algorithm as the NFL reflectance map.

### 2.3.5 Statistical Analysis of NFL Reflectance Loss Patterns

We tested whether most NFL reflectance loss patterns were consistent with known characteristic patterns of glaucoma. To perform this analysis, we categorized the loss pattern into five types: diffuse, wedge, other grouping, isolated, or none (Fig. 2). Diffuse pattern (full-width defect spanning more than a quadrant of the annular analytic area) would be consistent with severe glaucoma. In contrast, wedge patterns (contiguous superpixels connecting the inner and outer edges of the annular analytical zone) would be consistent with mild or moderate glaucoma where the damage was focal. Isolated (1 or more contiguous superpixels in a non-wedge configuration) could indicate measurement noise or mild disease. If two or more patterns were observed in the same eye, the one corresponding to a more severe glaucoma category was applied.

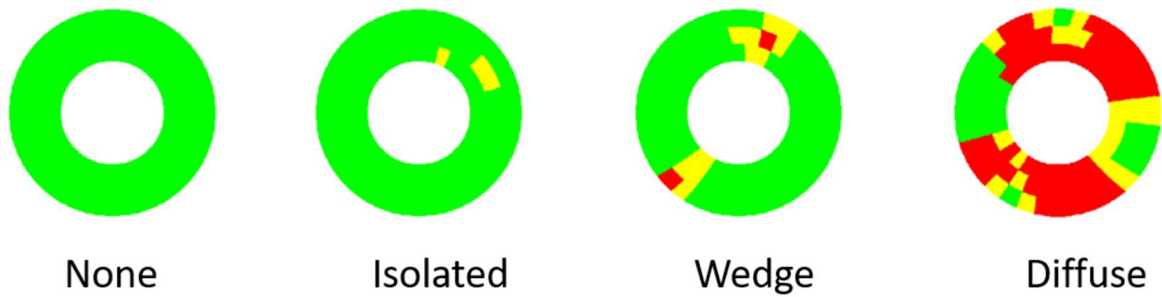

**Figure 2.** Four types of NFL reflectance loss patterns.

The two-sided Wilcoxon rank-sum test was used to compare the difference between the normal and glaucoma sub-groups. The diagnostic accuracy was evaluated by the area under receiving characteristic operating curve (AROC)[9] and by the sensitivity at the 99% specificity. The 95% and 99% specificity cutoff was estimated from normal eyes using the bootstrap method.[25] To account for inter-eye correlation, the AROC was computed based on the formula of Obuchowski,[26] which extended the nonparametric method of Delong et al.[27]

The sensitivity was compared using McNemar's test. For all parameters, age adjustment was applied to obtain an equivalent value at a reference age of 50 years.[28] Pearson correlation coefficients were calculated among NFL parameters and VF MDs. The coefficients were compared using the bootstrap method.[29] All analyses were done in Matlab R2021a with the statistics toolbox.

## 3 Results

### 3.1 Characteristics of the Study Participants

One hundred and five normal, 330 PPG, 165 early PG, and 73 moderate+ PG eyes were included in this study. Patients in all glaucoma sub-groups were older, had longer axial lengths, thinner central cornea thickness (CCT), worse VF MDs, and worse VF pattern standard deviations (PSD) than normal patients. Most of them were significant (p<0.05, Table 1) with the exception of VF MD and PSD in PPG group.

**Table 1. Characteristics of the Study Subjects**

|  | Normal | PPG | Early PG | Moderate PG+ |
|---|---|---|---|---|
| Participants# | 53 | 196 | 132 | 59 |
| Eyes# | 105 | 330 | 165 | 73 |
| Male/Female | 38/67 | 134/196 | 63/102 | 32/41 |
| age (Yrs) | 58.9±9.9 | 62.7±8.8* | 64.3±9.4* | 62.1±9.5* |
| Axial Length (mm) | 23.6±1.0 | 24.3±1.3* | 24.3±1.3* | 24.6±1.4* |
| IOP (mmHg) | 13.9±2.7 | 16.7±4.7* | 18.0±7.2* | 18.6±8.5* |
| CCT (μm) | 563.4±30.9 | 554.3±35.2* | 546.8±37.4* | 539.4±35.8* |
| VF MD (dB) | -0.0±0.9 | -0.4±1.4 | -2.3±1.7* | -11.0±4.0* |
| VF PSD (dB) | 1.5±0.4 | 1.7±0.6 | 3.6±2.4* | 11.1±2.9* |

*PPG = pre-perimetric glaucoma. PG = perimetric glaucoma. Moderate PG+: moderate and advanced PG. IOP=Intraocular pressure. CCT=central cornea thickness. VF MD = visual field mean deviation. VF PSD = visual field pattern standard deviation. Values for continuous variables are means ± standard deviations. \* denotes statistical significance of p < 0.05 compared to the normal group.*

### 3.2 Reflectance Patterns in Normal and Glaucoma Groups

The NFL reflectance map showed the highest reflectance in the inferotemporal and superotemporal regions (Figure 3). The standard deviation (SD) map showed slightly higher variability in the inner circles and the lowest variability in the inferotemporal and superonasal regions, with an average SD of 1.8 dB, and the peak SD of 2.4 dB. In glaucomatous eyes, the map showed reflectance loss most severe affected in the inferior area, followed by the superior area. Reflectance loss was correlated with the severity of glaucoma stages. The average loss was -1.3 dB in PPG, 2.7 dB in early PG, and -4.9 dB in moderate+ PG. The peak loss (i.e. inferior area) was -2.3 dB in PPG, -4.3 dB in early PG, and -8.3 dB in moderate+ PG eyes.

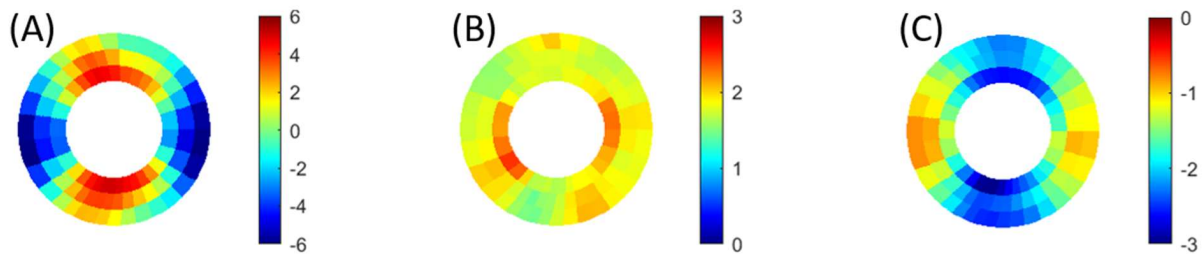

*Figure 3. NFL reflectance map (unit:dB) in super-grid for reference, normal and glaucoma groups. Maps of left eyes were mirrored before averaging. (A) average of reference group; (B) standard deviation of reference group; (C) average loss pattern of glaucoma group.*

Reflectance loss patterns correlated with glaucoma stages as well. More than half of PPG eyes (54%), 80% of early PG eyes, and 97% of moderate PG eyes exhibited either diffuse or wedge reflectance loss patterns. Seventy-seven percent of normal eyes exhibited isolated patterns, suggesting that these loss patterns were not diagnostic of glaucoma. Overall, 67% of glaucomatous eyes (PG and PPG) exhibited wedge-shaped or diffuse reflectance defect, compared to only 20% normal eyes exhibiting these patterns (p<0.001, Chi-square test). (Table 5)

**Table 2. Loss Pattern vs. subgroup**

|  | Normal | PPG | Early_PG | Moderate_PG+ |
| --- | --- | --- | --- | --- |
| **No** | 47(45%) | 63(19%) | 12(7%) | 0(0%) |
| **Isolated** | 37(36%) | 90(27%) | 21(12%) | 2(3%) |
| **Wedge** | 19(18%) | 105(32%) | 52(32%) | 9(12%) |
| **Diffuse** | 2(2%) | 72(22%) | 80(48%) | 62(85%) |

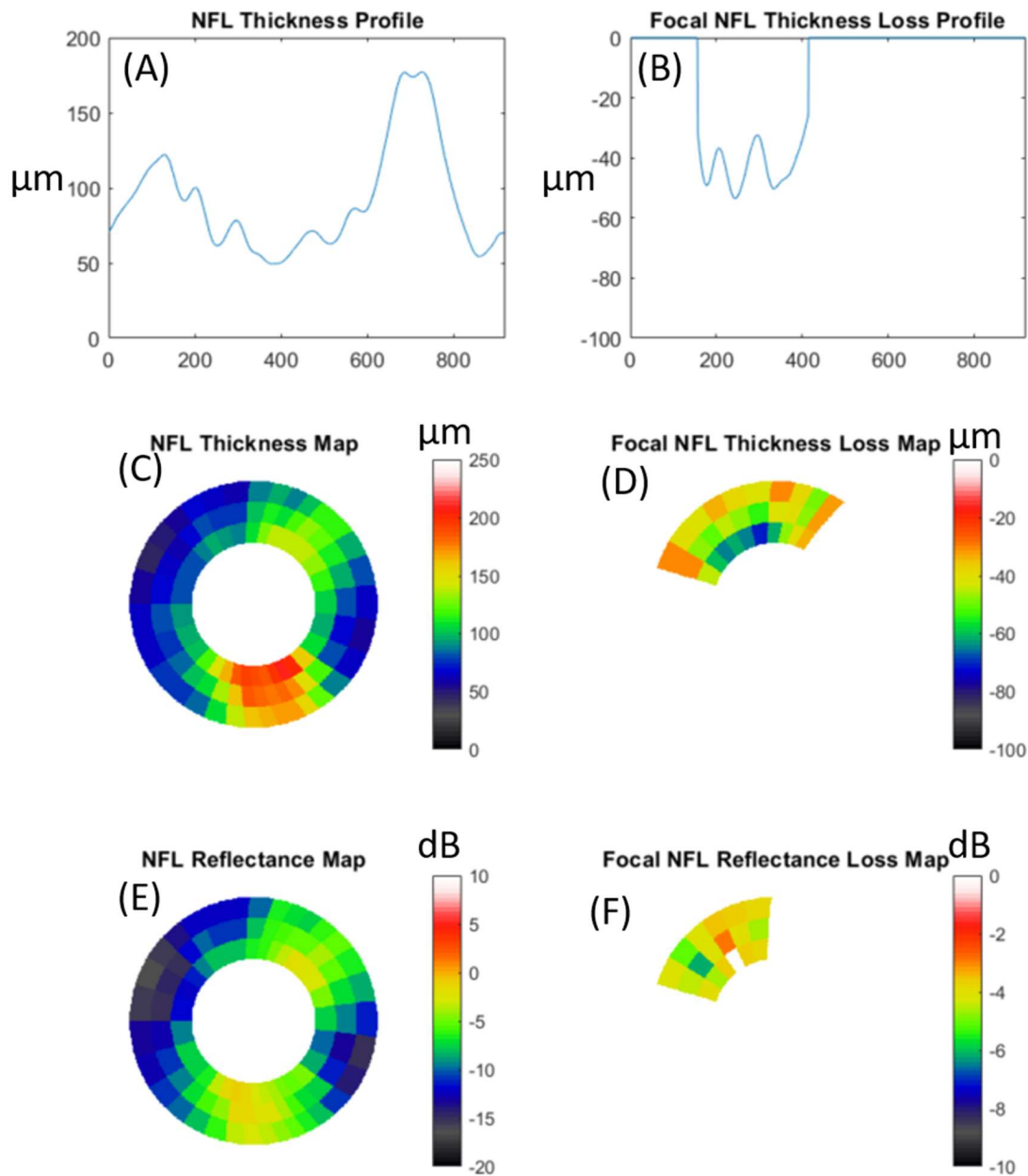

*Figure 4*. *Example of NFL thickness and reflectance from a glaucoma eye (A) NFL thickness profile at D=3.4mm; (B) focal loss of NFL thickness profile; (C) NFL thickness map at D=2.1~4.2mm; (D) focal loss of NFL thickness map; (E) NFL reflectance map at D=2.1~4.2mm; (F) focal Loss of NFL reflectance map. Note focal loss is set to 0 at area without significant NFL loss.*

3.3   Characteristic of Nerve Fiber Layer Parameters

Six parameter were averaged on NFL thickness and reflectance maps. (Figure 4) All NFL parameters were significantly different between the normal and glaucoma groups (Table 3). NFL average thickness decreased and NFL thickness focal loss increased with the stages of glaucoma. NFL reflectance also decreased and focal reflectance loss increased with the stages of glaucoma. Note that focal loss value was between 0 and infinity; a larger focal loss meant more loss than the reference values.

**Table 3. Group Statistics for Nerve Fiber Layer Parameters**

|  |  | Normal | PPG | Early PG | Moderate PG+ |
|---|---|---|---|---|---|
| Thickness Profile(μm) | Average | 98.7±9.1 | 93.4±10.5* | 83.9±12.4* | 76.4±14.0* |
|  | Focal Loss | 1.64±2.00 | 3.17±3.37* | 6.01±4.30* | 10.90±4.61* |
| Thickness Map(μm) | Average | 121.6±10.5 | 114.7±12.6* | 103.1±14.7* | 93.2±16.7* |
|  | Focal Loss | 1.94±2.82 | 5.60±7.57* | 14.42±12.30* | 25.95±15.10* |
| Reflectance Map(dB) | Average | -0.01±1.20 | -1.16±1.63* | -2.43±1.87* | -4.44±2.06* |
|  | Focal Loss | 0.24±0.45 | 0.95±1.10* | 2.04±1.71* | 4.15±2.11* |

*PPG = pre-perimetric glaucoma. PG = perimetric glaucoma. Moderate PG+: Moderate and Advanced PG. Values for continuous variables are means ± standard deviations. *, p-value < 0.001 compared to the normal group.*

### 3.4 Diagnostic Accuracy

Focal reflectance loss had significantly higher AROC (0.80, p=0.004) than the best NFL thickness parameter--the focal thickness loss from map (AROC=0.75). Focal reflectance loss also had the highest AROC in each glaucoma sub-group, but only significantly better than the best NFL parameters in PPG sub-group (p=0.009). Focal reflectance loss had higher AROC than the average reflectance, but the differences were not significant.

**Table 4. Diagnostic Accuracy (AROC) of Nerve Fiber Layer Parameters**

|  |  | PPG | Early PG | Moderate PG+ | Overvall Glaucoma |
|---|---|---|---|---|---|
| Thickness Profile | Average | 0.64±0.03 | 0.83±0.02 | 0.91±0.02 | 0.73±0.02 |
|  | Focal Loss | 0.62±0.03 | 0.81±0.03 | 0.96±0.01* | 0.72±0.02 |
| Thickness Map | Average | 0.65±0.03 | 0.85±0.02* | 0.92±0.02* | 0.74±0.02* |
|  | Focal Loss | 0.65±0.03 | 0.87±0.02* | 0.97±0.01* | 0.75±0.02 |
| Reflectance Map | Average | 0.71±0.03*+ | 0.86±0.02 | 0.96±0.02* | 0.79±0.02* |
|  | Focal Loss | 0.72±0.02*+ | 0.88±0.02* | 0.98±0.01* | 0.80±0.02*+ |

*AROC= area under operator receiver curve. PPG = pre-perimetric glaucoma. PG = perimetric glaucoma. Moderate PG+: Moderate and Advanced PG. Values are Estimated AROC ± standard error. *, p-value < 0.05 compared to the thickness Profile average. +, p-value < 0.05 compare to best NFL thickness parameter*

When combing PPG and PG as the overall glaucoma group, focal reflectance loss (p<0.028) had significantly higher glaucoma diagnostic sensitivity (0.37 and 0.55) than the best thickness parameters (0.33 and 0.50) when the specificity was fixed at 99% or 95% (Table 5). However, when the groups are separately analyzed individually (PG, early PG, and moderate PG), focal reflectance loss had the highest sensitivity but was not statistically significant compared to the best NFL thickness parameter.

**Table 5. Glaucoma Diagnostic Sensitivities.**

|  |  | 95% specificity | | | | 99% specificity | | | |
|---|---|---|---|---|---|---|---|---|---|
|  |  | PPG | Early PG | Moderate+ PG | Overvall Glaucoma | PPG | Early PG | Moderate + PG | Overvall Glaucoma |
| Thickness Profile | AVG | 0.26 | 0.61 | 0.77 | 0.43 | 0.12 | 0.40 | 0.66 | 0.27 |
| | Focal Loss | 0.23 | 0.53 | 0.93* | 0.41 | 0.07 | 0.25 | 0.6 | 0.19 |
| Thickness Map | AVG | 0.30* | 0.64 | 0.78 | 0.46* | 0.14 | 0.46* | 0.7 | 0.31* |
| | Focal Loss | 0.32* | 0.66 | 0.95* | 0.50* | 0.15 | 0.47* | 0.82* | 0.33* |
| Reflectance Map | AVG | 0.39* | 0.65 | 0.90* | 0.53* | 0.19* | 0.46* | 0.82* | 0.35* |
| | Focal Loss | 0.40* | 0.69 | 0.95* | 0.55*+ | 0.19* | 0.52* | 0.84* | 0.37*+ |

*AVG=Average. PPG = pre-perimetric glaucoma. PG = perimetric glaucoma. Moderate PG+: Moderate and Advanced PG. Values are Estimated AROC ± standard error. \*: p-value < 0.05 compared to the thickness profile average. +: p-value < 0.05 compare to best NFL thickness parameter*

### 3.5 Correlation with Visual Field

All NFL parameters had moderate Pearson correlation with VF MD (Pearson *r* between 0.43 and 0.64, Table 6), in which focal reflectance loss had the highest correlation (*r* = 0.64) and was significantly higher than all NFL thicknesses parameters (*p<0.001*). The average reflectance had a higher correlation than the average NFL thickness profile or map (p<0.001). Focal losses had higher correlations than averages (p<0.001). The NFL reflectance parameters highly correlated with NFL thickness (*r* =0.72 to 0.81), with the exception of moderate correlation with focal NFL loss from thickness profile (0.64 to 0.68). All correlations were statistically significant (p < 0.001).

We observed that the slope of average NFL parameters flattens as VF MD progresses, signifying a well-known "floor effect" when the NFL does not decrease further with worsening glaucoma in its later stages (Figure 5A, C, E). The floor effect became less apparent in focal reflectance loss. Piecewise linear regression was used to estimate the slopes in early and later stages. For example, the slope in reflectance, or the rate of reflectance change, was 0.119 at later stage (VF MD<-6dB), compared to 0.185 at earlier stage (VF MD>-6dB). In order to compare trends in different scales, we used the standardized slope, which is the slope normalized by the population standard deviation. The rate of focal reflectance loss change (0.153) is larger than the NFL thickness (0.102~0.145) in later stages of glaucoma. (Figure 5)

**Table 6. Pearson correlation with Visual Field**

|  |  | Reflectance Map | | Visual Field |
|---|---|---|---|---|
|  |  | AVG | Focal Loss | MD |
| **Thickness Profile** | AVG | 0.764 | 0.722 | 0.426 |
|  | Focal Loss | 0.641 | 0.681 | 0.569 |
| **Thickness Map** | AVG | 0.759 | 0.726 | 0.456 |
|  | Focal Loss | 0.748 | 0.808 | 0.568 |
| **Reflectance Map** | AVG |  | 0.929 | 0.565 |

| | |
|---|---|
| Focal Loss | 0.643 |

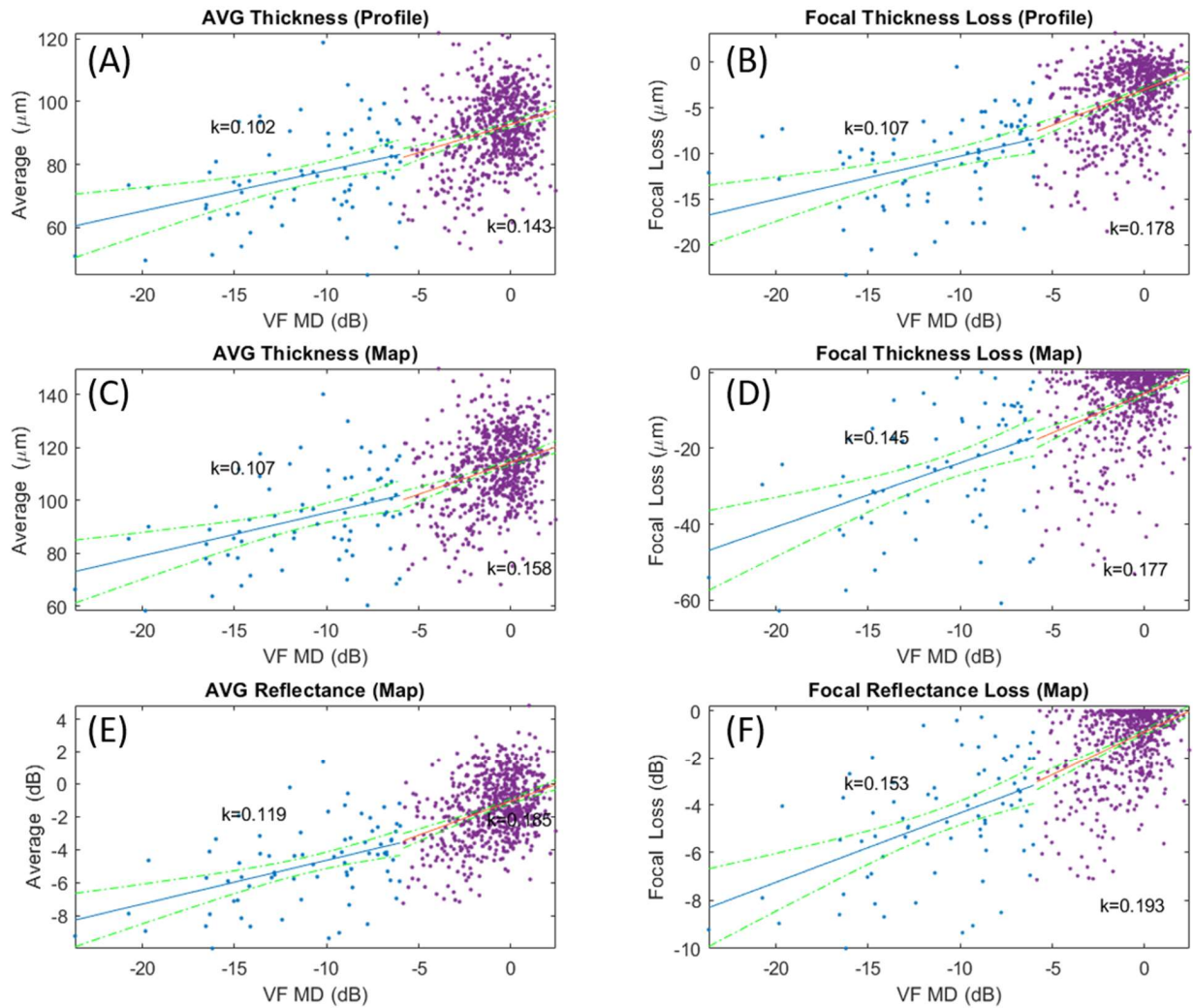

*Figure 5*. The association of nerve fiber layer (NFL) vs visual field (VF) mean deviation (MD). (A-B) Average and focal loss of thickness Profile (C-D) Average and focal loss of NFL thickness map (E-F) Average and focal Loss reflectance map. We divided data points into early stage group (VF MD>-6dB, purple points) and later stage group (VF MD<-6dB, blue points), and estimated the standardized slope k using linear regression for each group; orange line is the fitted line for early stage and blue line is for later stage; the green dash lines are the 95% confidential interval of fitted lines.

## 4  Discussion

In this study, we compared the glaucoma diagnostic performance of focal NFL reflectance loss to NFL thickness on circular peripapillary scans of a 26kHz older version of SD-OCT. The result confirmed previous finding that focal NFL reflectance loss could significantly improve the diagnostic accuracy, especially in pre-perimetric glaucoma. In addition, we observed NFL reflectance had a higher correlation with the visual field parameter. We also noticed that the NFL reflectance was OCT system depended.

Comparing the AROC and sensitivity between focal NFL reflectance and the best NFL thickness parameter, this study showed that focal NFL reflectance loss improved diagnostic accuracy in all glaucoma sub-groups, but only significantly in PPG or all glaucoma eyes. This finding agreed with Liu's study, in which they found that AROC of NFL reflectance was higher than NFL thickness, but only significantly in glaucoma suspect.[30] In a previous study using a newer OCT model, we found a significant difference in sensitivity between NFL reflectance and the best NFL thickness parameter for PPG and PG.[19] These studies, together with the result from this study, confirmed that focal NFL reflectance analysis could improve the diagnostic accuracy of NFL.

The sensitivity of NFL reflectance is 19% for PPG and 52% for early PG in this study. It is possible to improve the sensitivity by combining reflectance loss with other OCT parameters, like cup-disc ratio, rim area, BMO-MRW, ganglion cell complex thickness map.[8, 10, 14-18] It may be useful to apply deep learning as a more robust analysis to account for interactions between variables.[31]

When correlating to VF, we found that NFL reflectance correlated better than thickness, and that focal loss better than average values. Therefore, focal NFL reflectance loss correlated best with VF MD glaucoma progression comparing to average NFL thickness. It has less "floor effect" than the average NFL thickness. This finding suggests that focal NFL reflectance loss may be suitable for monitoring glaucoma progression over a broader range of glaucoma severity. We will test this hypothesis using the longitudinal AIG data, which permits analysis over 7-11 visits.[6, 32]

The NFL reflectance was associated with age and axial length in normal eyes. Surprisingly, we found significant differences between clinic sites, even after adjustment for age and axial length. The inter-site variation in the ethnic composition may be one of the reason,[20] but we did not find significantly different NFL reflectance among races in the normal reference in AIG study. Since the 3 sites used the same OCT models, the difference may be also due to the operators or the tuning of the systems. Potential system factors included the polarization state of the OCT beam in the sample arm and the signal-to-noise ratio (SNR) of the system (measured with a standard sample reflector).[33] Potential operator factors included the positioning of the OCT system relative to the eye, which affected the incidence angle of the OCT beam on the retina and the focusing of the OCT beam. These factors may limit using NFL reflectance as a diagnostic parameter. Newer OCT systems generally employ greater automation and better alignment and focusing aids, reducing the inter-device and inter-operator variability.

This paper's diagnostic performance of focal NFL reflectance loss is not as high as our original publication on this new approach.[19] This paper analyzes an older (and bigger) dataset acquired using an slower, previous generation OCT system that sparsely samples the peripapillary retina, compared to the full volumetric imaging used in our previous paper. The much higher proportion of normal control eyes with reflectance defects is notable, which could be due to less consistent focusing and lower SNR of the older system. The sparse sampling of the older scan pattern also led to less precision in centering the NFL maps on the optic disc boundary. Despite these limitation, this larger and older dataset was able to confirm the advantage of focal NFL reflectance loss as a diagnostic parameter over the average NFL thickness, which is the current standard of care. NFL reflectance is likely a useful diagnostic and monitor tool using newer, faster OCT systems capable of volumetric scanning of the optic disc and surrounding retina.

# 5 Conclusions

We have confirmed that focal NFL reflectance loss analysis significantly improves diagnostic accuracy than the widely used NFL thickness parameter with a large dataset.

Author Contributions

O.T., D.C. and D.H. designed the study. O.T. and D. C. and D. H. wrote the manuscript. O.T. supervised the project. D.C. directed the statistical analysis, DG, BF,RV and JS directed the data collection in clinical sites and revised the manuscript. AC revised the manuscript.

Competing Interests Statement

OHSU, Dr. Tan, Dr. Huang have a significant financial interest in Optovue, Inc., a company that may have a commercial interest in the results of this research and technology. These potential conflicts of interest have been reviewed and managed by OHSU.

Acknowledge

This study was supported by NIH grants R21 EY032146, R01 EY023285, P30 EY010572, and an unrestricted grant from Research to Prevent Blindness to Casey Eye Institute.

32.     Zhang X, Francis BA, Dastiridou A, et al. Longitudinal and Cross-Sectional Analyses of Age Effects on Retinal Nerve Fiber Layer and Ganglion Cell Complex Thickness by Fourier-Domain OCT. *Transl Vis Sci Technol* 2016;5:1.

33.     Choma M, Sarunic M, Yang C, Izatt J. Sensitivity advantage of swept source and Fourier domain optical coherence tomography. *Opt Express* 2003;11:2183-2189.
18